\documentclass{llncs}
\usepackage{caption}
\usepackage [english]{babel}
\usepackage [autostyle, english = american]{csquotes}
\DeclareCaptionType{copyrightbox}
\MakeOuterQuote{"}
\usepackage{graphicx}
  \usepackage{times} 
\begin{document}

\title{Doris: A tool for interactive exploration of historic corpora (Extended Version)}
\author{Sreya Guha}
\institute{Castilleja High School}

\maketitle

\begin{abstract}
Insights into social phenomenon can be gleaned from trends and patterns in corpora of documents  associated with that phenomenon. Recent years have witnessed the use of computational techniques, mostly based on keywords, to analyze large corpora for these purposes. In this paper, we extend these techniques to incorporate semantic features. We introduce Doris, an interactive exploration tool that combines  semantic features  with information retrieval techniques to enable exploration of document corpora corresponding to the social phenomenon. We discuss the semantic techniques and describe an implementation on a corpus of United States (US) presidential speeches. We illustrate, with examples, how the ability to combine syntactic and semantic features in a visualization helps researchers more easily gain insights into the underlying phenomenon.

\end{abstract}

\section{Introduction}

One way of understanding social phenomenon or the behavior of groups is to analyze the discourse within the group. Such a discourse is often scattered across different documents in a large corpus. In any given document corpora, analysis of words, topics and evolution of the discourse provide deeper insights into the social phenomenon. This is especially true for historic analysis of any social phenomenona.

In recent years, computational tools have been developed to analyze large corpora of documents. A notable application of the use of computational techniques was the authentication of the authorship of the works attributed to Shakespeare. Computational techniques have enabled a deeper analysis into the playwright's work to uncover if he was the sole author \cite{PennNews}.

Another example of the use of computational techniques was in determining the likely authorship of the Federalist Papers. The Federalist Papers, published between 1787 and 1788, are a collection of 85 essay and letters written to promote the ratification of the United States Constitution. While published under an alias, the authors were identified to be Alexander Hamilton, James Madison and John Jay. However, the authorship of 12 papers are disputed. While Hamilton had claimed credit for those disputed papers, analyzing the discourse has introduced the possibility that it was Madison who wrote these papers \cite{constitutionfacts.com}. 

Another area of development is the use of interactive user interfaces that enable a researcher to explore a corpus of documents. The ability to visualize specific words in the context of a corpus can help researchers gain insights into the evolution of the discourse on particular topics. The recent work by Schmidt et. al. 
\cite{schmidt2012words} has illustrated the power of interactive tools for deeper understanding of a corpus of speeches, such as those by US Presidents. 

Most of the tools currently in use for researching social phenomenon through associated document corpora perform a relatively shallow analysis. As we discuss in the next section, much of this work is focused on the distributional properties of words and phrases in these texts. These tools can be applied to a new corpus without any manual pre-processing as they tools are driven only by the words in the document.  This generality,  however, can also be a shortcoming. Many interesting questions cannot be expressed purely in terms of the words that appear in the documents. 

Consider, for example, the corpus of US Presidential State of the Union speeches (a State of the Union speech is an annual speech given by the President of the United States, to the United States Congress). Through the course of its history, the role and status of Native Americans has been an issue, to varying degrees. In the early years of the US, Native American relations were talked about more frequently than they are today. However, it is difficult to perform this analysis by looking at the distribution of a word or phrase. Difficulties arise from the fact that the terminology used to refer to them has evolved over time. Not only are they referred to now as Native Americans, as opposed to 'Indians' or 'Red Indians' a hundred years ago, but also we find that two hundred years ago, many of the references to them were specific tribes (Apache, Cherokee, etc.) and  legislation. Ideally, we would like a model that provides "Native Americans" as a first class model feature, allowing us to slice and dice by this features, just as we could using a word or phrase.

We have seen the emergence of machine learning techniques (\cite{mikolov}, \cite{blei2003latent}, \cite{glove}) to build deeper semantic models. Such models are not interactive --- they run offline but can sometimes give insights into the social discourse. For example, Evans \cite{soutopicmodels}  uses topic models to show the rise of partisanship in US Presidential state of union speeches.

In this paper, we describe Doris\footnote{Named after  Doris Goodwin, noted US presidential historian.}, an interactive exploration tool that combines the generality of keyword based approaches with the deeper semantic understanding enabled by both by Semantic Web markup and by machine learned models. 

Our tool (available at http://pres-search.appspot.com) allows the user to search the corpus by specific word, phrase or topic and see the distribution of mentions across Presidents. The user can also restrict the search by president or type of speech (such as State of the Union, Proclamation and Executive Action). This feature allows the researcher to look at how a president has chosen to display and present his agenda in regards to a specific topic. The search results are accompanied by graphs plotting the distribution of documents satisfying the search criteria. These graphs enable researchers to gain insights into the evolution of the discourse, as captured by that search, over time.

Our long-term goal is to enable interactive exploration of large document corpora at a semantic level. In this paper, Doris combines categorical filtering based on semantic categories together with classical keyword based search to create a tool that can help 
explore a corpus of documents. We apply this tool to a corpus of over 12,000 documents of US Presidential statements,  including the State of the Union speeches, Proclamations, Inaugural addresses and Executive Actions. We propose a Schema.org compliant vocabulary to capture the various aspects of the metadata  associated with this class of political discourse. We describe a tool we have built that combines semantic markup data provided in this vocabulary together with the text of documents to create a search tool.  We use this tool to explore this document space and illustrate the power of the tool with some conclusions that follow.


\section{Related Work}


  Though the use of computational techniques for analyzing text corpora in the context of social science research is relatively new, there is already a rich and growing body of work that use various techniques drawn from the information retrieval community in order to better understand social and political phenomenon.The work by Shen, Aiden, Norvig, et. al. \cite{michel2011quantitative}, which performed a quantitative analysis of the unigrams and bigrams in millions of digitized books, though very simple in its analysis, was very influential. It illustrated how even simple techniques, when applied across very large corpora, can provide interesting insights.

Ben Schmidt \cite{benschmidtsotu} uses the bookworm database to visualize the occurrence of certain words in the State of the Unions by American presidents. Given a particular State of the Union, the user can choose certain words and see a distribution throughout all presidents. 

 Baker, Gabrielatos, et. al \cite{doi:10.1177/0957926508088962} examine a 140 million word corpus of British news articles concerning refugees and immigration using techniques usually associated with corpus linguistics. They study the extent to which methods normally associated with corpus linguistics can be effectively used by critical discourse analysts.

In \cite{grimmer_2016}, Grimmer uses a statistical topic model on press releases from the House of Representatives from 2005 to 2010 to demonstrate the shift in portrayed representation due to electoral pressure. The author shows how members of the House change rhetoric, specifically in terms of taking credit, due to political pressure. 

Hillard, Purpura and Wilkerson \cite{doi:10.1080/19331680801975367} examine over 300,000 congressional bill titles (that researchers have assigned topics to) and use supervised learning algorithms to allocate topics. The authors show a successful method of classifying large sets of data computationally.
Hopkins and King \cite{AJPS:AJPS428} develop a method that gives estimates of category proportions for large sets of data. Using data sets that include relevant political opinions, the authors focus on document category proportions rather than absolute counts of individual categories. 

Laver, Benoit, et. al. \cite{laver_benoit_garry_2003} presents a unique way of determining political stances using computational techniques based on language-blind scoring technique. The authors introduce uncertainty measures, allowing researchers the ability to make better observations. 
Thomas and Pang \cite{Thomas:2006:GOV:1610075.1610122} use a corpus of U.S. congressional floor debates to attempt to determine support or opposition in certain issues. The authors take into account Support Vector Machines and the fact that the data is conversational to create a classification framework.



In addition to the work involving the use of computational techniques in the social sciences, we draw on work on Topic Models \cite{blei2003latent} and word embeddings \cite{mikolov}. 
Our interface is influenced by the work of Freeman and Gelernter \cite{lifestreams}, in which they introduced the idea of temporal presentation of a set of documents. Bergman, Beyth-Marom, et. al.  \cite{bergman2008improved} adapted this work to the context of search interfaces. Recent years have seen the adoption of this class of interfaces in widely used software systems, including the Apple Mac interface.

\section{Methodology}

Our goal is to create a tool for interactive exploration of a corpus of documents that captures the discourse in some social phenomenon. With any large corpus, we need the ability to begin the exploration from different starting points. Keyword based search offers a good metaphor for this. We augment the `raw' text of the documents with structured/semantic data, including metadata (author, date, etc.) and annotations that capture higher level semantics of the topics discussed in these documents.

We apply our tool to a corpus of documents from US Presidency Project at the University of California, Santa Barbara, US Presidency project \cite{ucsb}, that includes all the US Presidential State of Union Speeches, Proclamations, Executive actions, Proclamations, etc., giving us a total of 12,345 documents. We gather the following kinds of metadata: type of speech, author (president) and date for each. Since, the metadata is not directly embedded into these documents (like it often is in web pages), in addition to the text corpus, our processing pipeline accepts files containing annotations  on these files expressed in RDF or RDFa. In addition to simple metadata about the texts, the kind of exploration we seek to enable benefits greatly from annotations that capture semantic aspects of documents. We now describe our work on each of two kinds of annotations.

\subsection{Metadata}
 We extracted the metadata from the Presidency websites\cite{ucsb} by using a set of scrapers. 
 Some of the vocabulary for expressing the metadata is already available in schemas such as those from Schema.org. Other aspects of the metadata, such as the kind of speech/document, are not part of any well-known vocabulary we know. In order to facilitate this, we have developed a number of vocabulary terms, which are in discussion, for inclusion into Schema.org.

 From existing Schema.org, we use the vocabulary items \\$datePublished$, $author$ and $title$. An important aspect of the documents in this corpus is the kind of document: State of Union, Proclamation, etc.  Schema.org has a very general class called 'CreativeWork' and we can introduce subclasses under this to represent these kinds of documents.  While it is easy to simply introduce four new types as subclasses of CreativeWork, it is clear that these are just four in a much large landscape of political documents. After a set of discussions involving the Schema.org community, we used the following vocabulary.
 
\subsubsection{Political Discourse Vocabulary}

We propose a number of additions to the existing vocabulary at Schema.org. Schema.org already
has the properties we require. We augment the existing vocabulary with the two classes: 1) subclasses of CreativeWork (e.g., PublicSpeech, PressRelease, Proclamation, ExecutiveAction), and 2), subclasses of Speech (e.g., InauguralAddress, CommencementAddress, CampaignSpeech, StateOfUnionReport). Though some of these items, such as $StateOfUnionReport$, are specific to the political structured of the United States, most of the new added terms apply not just to US politics, but more generally, to any political discourse.

\subsection{Semantic Annotations}
As discussed in the introduction, simple word level treatments are not capable of capturing trends that involve the significant, correlated variations in vocabulary (such as the language around partisan issues such as abortion and gun control), the evolution of vocabulary (such as the words used to refer to people of African origin) or the differences in granularity (such as Cherokee vs Native American Tribes). A number of different techniques have been developed over the years to extract higher level or more semantic abstractions, or topics, of documents. Since we would like to be able to use different techniques, we accept any number of annotation files, each of which can be generated by different tools. Each annotation file typically contains a number of 'schema:about' statements about one or more of the documents, each associating a document with one of the `topics' covered in the document. We now describe the mechanisms used for generating topics for the US Presidency corpus.

Since the only raw features that are available  are typically words, clusters of words, either weighted or unweighted, are the most basic way of modeling topics. 
Each topic can be characterized by the mention of one or (preferably) more of a set of concepts, each of which in turn can be referred to by a set of alternative words phrases. Or goal is to bootstrap to a comprehensive set of words/phrases for each topic, with as little manual work as possible.
We start with a very simple keyword cluster mechanism and use a variety of techniques to enhance these clusters. 

\begin{figure} [!ht]
\centering
\includegraphics[scale=.5]{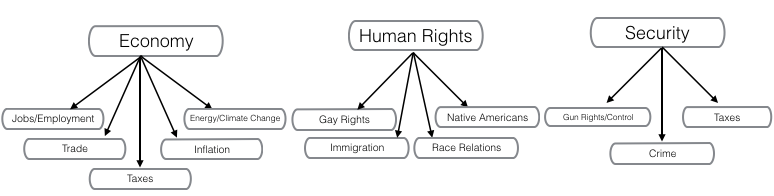}
\caption{A portion of the hierarchy of topics used. Note that some of the items actually fall under multiple parents, but for the sake of readability, the additional arcs are omitted from this picture. E.g., Climate change is parts of not just economy but also International and Health and arguably even Human Rights. Similarly, Terrorism is also part of Security.}
\label{topics}
\end{figure}
In this work, we used a taxonomy of topics from earlier work \cite{madhusreya}, which is shown in figure \ref{topics}. Note that this is not a strict taxonomy, since some of the nodes, such as `Terror' fall under multiple categories. Though the taxonomy used in the current version of the system is relatively small, the general methods described here apply to much larger taxonomies. We  began with a manually generated, small (4-8) keywords (positive and negative) for each topic. We extended each of these sets of keywords using the following techniques.

\subsubsection{Word co-occurrence:} Co-occurrence, i.e.,  the above-chance frequent occurrence of two terms from a text corpus alongside each other, can be interpreted as an indicator of semantic proximity. That is, occurrence of the co-occurring word can also be seen as a signal of the piece of text referring to that topic. We computed co-occurrence of terms at the sentence level and generated a list of the most frequently occurring words, above a threshold, for each manually added keyword and added these to our list of keywords for each topic. 

\subsubsection{Word embeddings:} Techniques developed by Mikolov \cite{mikolov} and many others has demonstrated how word embeddings can capture rich semantic meaning in a way that traditional bag-of-words models cannot. By constructing models to predict a word from its context (or vice versa), these models allow us to map words/phrases to vectors. Most notably, words that are "close" to each other in the vector space are likely to share similar contexts (and thus meaning). We use the Glove \cite{glove} prebuilt models in this work here. For each of the keywords corresponding to a topic, we collect the terms above a threshold of similarity and add them to the keyword list for that topic. 

\subsubsection{Topic modeling:} Starting with Blei, Jordan, et. al.\cite{blei2003latent}, there has been substantial work on identifying a set of `topics' which can be combined in different proportions to generate the articles (modeled as a bag of words) in a corpus. Topic modeling has become an effective tool for the discovery of underlying semantic themes in document corpora. Consequently, there are several widely used packages that can be used to generate topics. In this work, we 
use the ldaModel class in the Gensim \cite{gensim} tool, which assumes that the documents are generated from a set of topics using a Latent Dirichlet Allocation (LDA). We generate a set of topics for the 
documents in the corpus. We examine these topics, identify those that correspond to our set of topics and import the top 3 words corresponding to the keyword set for that topic. The number of topics to be generated is an important input to the LDA algorithm. After some experimentation, we found that 300 topics yielded the most number of topics corresponding to our topic list.

We start with a small manually generated list of keywords for each of the topics in figure \ref{topics} (every leaf node is assigned a set of keywords, which are inherited by their parents, which might be given an additional set of keywords). The keywords may include phrases to be interpreted as a bag of words, phrases that should be matched exactly, negative keywords, etc. We augment these initial keywords with additional keywords by using the techniques listed above.
We run the final set of keywords against the corpus to generate a list of topics that each document covers. This data is output into an annotations file, which together with a file containing the metadata (speakers, speech type, date) is consumed by the search engine run time.

\begin{figure}
  \includegraphics[scale=.4]{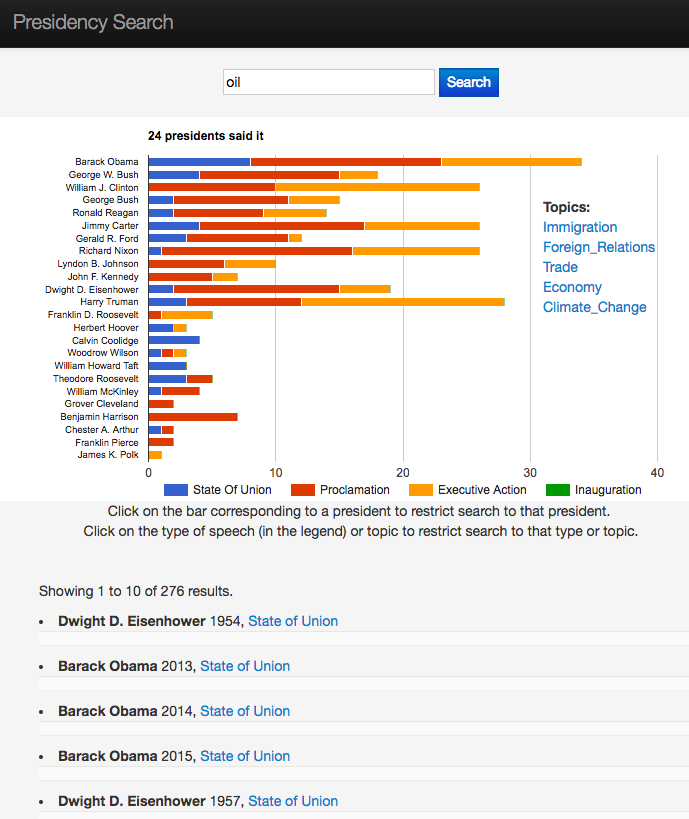}
  \caption{Results page for the query `oil'. The chart at the top gives the distribution of results, clustered by presidents, who are sorted in temporal order. The bar for each president is composed of segments corresponding to the type of document. On the right are the most frequently mentioned topics in the documents retrieved. The results can be restricted by clicking on the type of speech (in the labels), the president or the topic.}
  \label{psearch1}
\end{figure}

\begin{figure}
  \includegraphics[scale=.4]{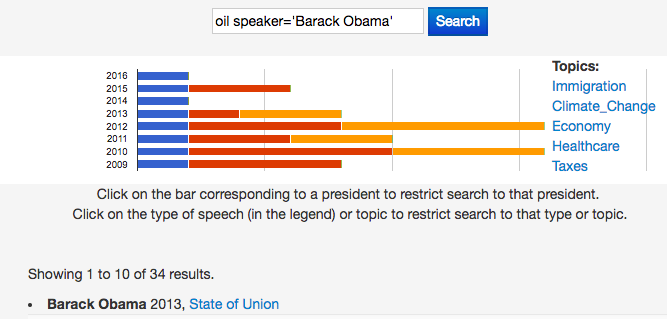}
  \caption{Results page for the query 'oil', restricted to Obama}
  \label{psearch2}
\end{figure}

\subsection{Exploration interface}
Doris is a hybrid between tools such as traditional search,  which is aimed at enabling the user to find the most relevant result, and Google's Ngram viewer \cite{ngram}, which is primarily focused on distributions of terms. We extend both the charting and search capabilities of those systems with the addition of semantic categories and metadata and their use in the interface. In this section, we describe some of the features of this tool and illustrate them with screen shots.

The user starts with a query, which could be a single word or a set of words. The results page, for a simple query `oil' is shown in Fig \ref{psearch1}. The bottom half of the page contains the first 10 results and is similar to a traditional search. The top of the page contains an interactive plot of the distribution of results. The results are collated by the speaker, which is plotted on the Y-axis (or optionally X-axis), arranged in a temporal order. Each bar is split into sections for the different kinds of speeches (State of the Union, Proclamation, etc.). On the right/bottom, we show the top 5 topics that appear in these results. 

\begin{figure}
  \includegraphics[scale=.5]{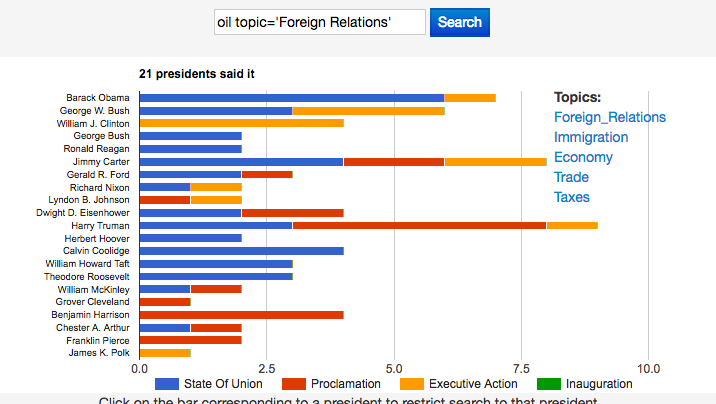}
  \caption{Results page for the query `oil', restricted to speeches covering the topic 'Foreign Relations'}
  \label{psearch4}
\end{figure}

Clicking on a topic restricts search to that topic (see figure \ref{psearch4}). Clicking on the bar corresponding to a president restricts attention to the speeches of that president. As shown in figure  \ref{psearch2}, in this view, we aggregate a single president's speeches by year. The set of search results, which is in the bottom half of the page, is also kept updated through this exploration. In a different view of the interface (see figure \ref{economy}), when the search is restricted to  a topic node that has children, the bars for each president correspond to the subtopics of that topic node.

\section{Analysis / Discussion}
 While it is possible to evaluate a tool such as the one presented here, using traditional
information retrieval metrics such as precision and recall, simply performing as well (or even slightly better) than current search tools would not justify such an effort. The primary goal of this tool is to make it easier for students and researchers to gain insights from large document corpora. In that spirit, we discuss some patterns that are apparent through our tool that are not obvious in traditional systems.
Given the huge increase in the number of Proclamations and Executive orders in the recent past, in order to normalize for comparisons, unless otherwise mentioned, we restrict our attention here to State of the Union addresses.

\begin{figure}[h]
\begin{center}
  \includegraphics[scale=.6]{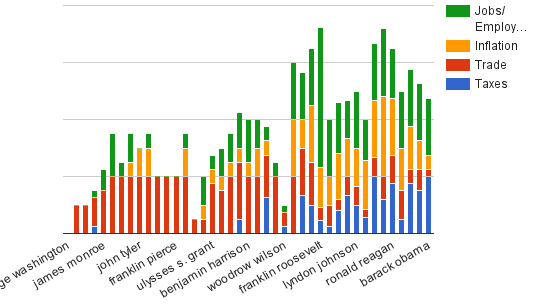}
  \caption{Results for the topic `Economy' in State of the Union addresses. This graph, unlike the earlier ones, is horizontal. Also since the topic 'Economy' has subtopics, the bar for each president is broken down by subtopics.}
  \label{economy}
\end{center}
\end{figure}

Figure 2: We can see the frequency of the word `oil' increasing over time, starting from Harry Truman's presidency. Furthermore, `oil' appears mainly in Proclamations and Executive Actions and less in State of the Unions.
Isolating oil and foreign relations together, like in figure 5, shows peaks in Harry Truman, Jimmy Carter and Barack Obama.

Doris allows us to observe such patterns, which are not evident through the keyword search tool at the UCSB presidential library site. Doris, unlike existing search tools, allows researchers to frame questions like why did `oil' gain prominence post World War II and why has `oil' featured more in Proclamations and Executive Actions and less in State of the Unions. 

Figure \ref{economy}:  While `Economy' is currently one of the biggest issues and is discussed frequently in State of the Unions, it was not as prominent with early presidents as seen in figure \ref{economy}.
The `Economy' started to gain prominence around the Great Depression, more specifically post Woodrow Wilson. Further, we can see from Figure \ref{economy}, in the early days of the United States, discussions of the `Economy' was mainly focused on trade. Since Franklin D. Roosevelt's presidency, the focus of the economy has shifted to jobs and employment. Both the rising prominence of the `Economy' and transition from `Trade Relations' to `Jobs and Employment' are seen through the visualization in figure \ref{economy}.  These trends, which are not as evident by either reading individual documents or traditional search engines, can help researchers identify noteworthy trends.

\begin{figure*}
\begin{center}
  \includegraphics[scale=.94]{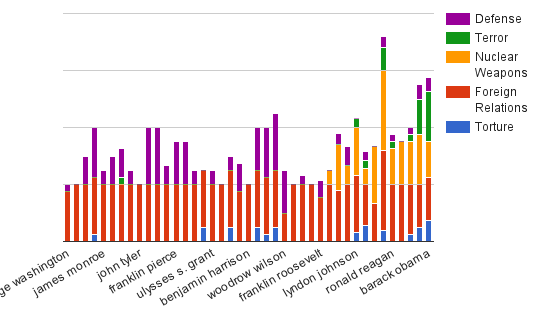}
  \caption{Results for the topic `Foreign Relations' in State of the Union addresses. Since this topic has subtopics, the bar for each president is broken down by subtopics.}
  \label{foreign_rel}
\end{center}
 \end{figure*}
This example illustrates the core strength of an interface that combines the ability to search across both words and topics, presenting them in an interactive graphical form. Semantic features (such as the topic `Economy') are vital to be able to perform this kind of analysis. This tool enables researchers to compare the evolution and relevance of certain topics such as how `Economy' has evolved compared to `Foreign Relations'.

Figure \ref{foreign_rel}: In contrast to the `Economy', we can see that `Foreign Relations' and `Defense' have remained a part of the State of the Union since its inception. `Nuclear Weapons' and `Terror' have recently emerged as important subtopics, gaining prominence in recent decades.

\begin{figure}
\begin{center}
  \includegraphics[scale=.5]{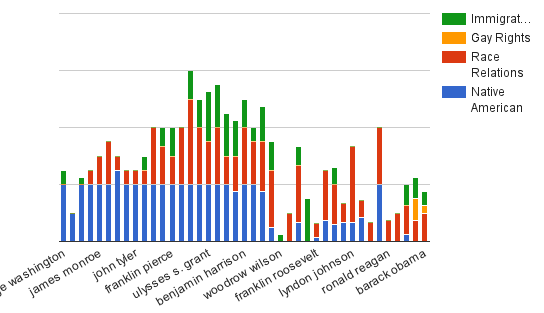}
  \caption{Results for the topic `Human Rights' in State of the Union addresses. Since this topic has subtopics, the bar for each president is broken down by subtopics.}
  \label{human_rights}
\end{center}
\end{figure}

 Figure \ref{human_rights}: In the 18th and 19th century, `Native Americans' were an important part of the discussion of `Human Rights' (or lack thereof). However, there is a recent fall in the discussion of `Native Americans' in State of the Union speeches (last mentioned briefly in President Clinton's State of the Union in 2000). `Gay rights' is a recent issue and only been incorporated into State of the Union speeches in the last few presidencies. 

We can see from figure \ref{human_rights} that `Race Relations', specifically the rights of African Americans, has always been a central part of the discourse on `Human Rights'. The issue gained prominence in the mid 1800's and we see a peak right before President Lincoln in President Buchanan's speeches. 

These observations are not evident without the help of the visualization in figure \ref{human_rights}. This tool allows researchers to delve deeper, ask important questions and find areas for further research. For example, after seeing a peak in President Buchanan's speeches and not Lincoln's speeches, a researcher might try to investigate the underlying cause.


\subsection{Future Directions}

 The goal of studying document corpora is ultimately to gain a better understanding of the underlying social phenomenon. We showed how a combination of traditional search and semantic annotations can make it easier to explore a document corpus and understand trends. There are a number of directions for future work, some of which we now discuss.
 
  Most of the attention in the version of Doris presented in this paper focuses on one dataset, i.e., the US Presidential dataset from the Presidency Project\cite{ucsb}. In the future, we hope to enhance this with many other datasets that are centered around US history. In addition to speeches by US Presidents, we plan to include documents from other branches of the government, including congress and the judicial system. 
    A number of trends/patterns in social phenomenon, especially historic phenomenon, are related to or triggered by events. There are many rich sources of structured data about various historic events, including Wikidata. Our goal is to be able to annotate portions of our graphs with relevant events. The key problem here is that of determining which events are most relevant. 
 
 The current version of Doris relies on human creation of the list of topics, the initial set of keywords for each topic and for identifying which of the automatically generated topics (from Topic Modeling) corresponds to the human created topics. We are looking into both automating some of these steps and into creating pre-built libraries of topics, e.g., using Wikipedia categories as topics.
Another option is to let users define new topics as they use the tool. As they use the system, if the different queries in a session are all different keywords associated with a topic they are trying to research, the system can define these as the seed for a new topic, pull in new keywords from Word2Vec, etc. Users should then be able to save these topics and  make them available to the larger research community.

 \section{Acknowledgements}
  I would like to thank Dr. Christy Story for her guidance and mentorship through this project. I would also like to thank Tasha Bergson-Michelson, Lillian Hoodes, Dave Lowell, Kyle Barriger and Ann Greyson.
 
 \bibliographystyle{unsrt}
\bibliography{flabitus}

\end{document}